\RequirePackage{ifpdf}
\documentclass[11pt,a4paper,notoc]{article}
\usepackage{jheppubme}

\newcommand{\ltsima}{$\; \buildrel < \over \sim \;$}
\newcommand{\simlt}{\lower.5ex\hbox{\ltsima}}
\newcommand{\gtsima}{$\; \buildrel > \over \sim \;$}
\newcommand{\simgt}{\lower.5ex\hbox{\gtsima}}
\newcommand{\gr}{\kern 2pt\hbox{}^\circ{\kern -2pt K}} 

\def\msun{\rm{\,M_{\odot}}}
\def\kms{\rm{\,km~s^{-1}}}
\def\pc{\rm{\,pc}}
\def\kpc{\rm{\,kpc}}
\def\mpc{\rm{\,Mpc}}
\def\ergs{\rm{\,ergs^{-1}}}

\def\sxu{\rm{\,cm^2}{gr^{-1}}}
\def\concI{\rm{c^{NW}_{200}}}
\def\concII{\rm{c^{SE}_{200}}}
\def\SDM{\rm{SZ-DM}}
\def\ctsn{\rm{\,counts~arcsec^{-2} }}
\def\dms{\rm{\,gr{\,cm^{-2}}}}
\def\rcp{r_c^{\rm{prm}}}
\def\XDM{\rm{X-DM}}
\def\BDM{\rm{BCG-DM}}

\def\BX{\rm{BCG-X}}
\def\lcdm{$\Lambda$CDM}
\def\om{\Omega_m}
\def\oml{\Omega_{\Lambda}}

\newcommand\gtilde{\stackrel{\sim}{\smash{g}\rule{0pt}{0.8ex}}}

\title{An SIDM simulation   of the merging  cluster El Gordo
 and  its tension between the post collision  DM density profiles
and weak lensing constraints,\footnote{
Talk presented at the  Valencia workshop on
 ``Small Scale Structure of the Universe and Self-Interacting Dark Matter'',
June 9-20, 2025, Valencia , Spain.
}}

\author{Riccardo Valdarnini,\footnote{Email:valda@sissa.it}\\ \it
SISSA, Via Bonomea 265, Trieste, Italy}

\abstract{
We review recent findings from a detailed simulation study of the 
 merging  cluster  El Gordo and  present new results inferred from weak lensing 
data.  We found that the observed spatial offsets between the different mass components 
are well reproduced in merging simulations that include 
self-interacting  dark matter (DM),  with an elastic cross-section per unit mass 
of approximately $\sigma_{DM}/{m_X}  \sim 4 \sxu$.
Moreover, a relative  line-of-sight  peculiar  velocity
  on the order of several hundred $\kms$  is found  
between the two stellar components  of the colliding subclusters.
These findings  strongly suggest the possibility that, in a very energetic cluster 
collision, DM could possess   collisional properties.

However, the  self-interacting DM merger model presented  here  is  not 
without difficulties. The values found for  $\sigma_{DM}/{m_X} $ being in conflict  
with the current upper bounds  on cluster scales. 
As a solution to this tension  we argue that in major cluster mergers   
the physical modeling of DM interactions,  based on the  scattering of DM particles,  
should be considered  too simplistic.

Additionally, the DM halos  of the post-collision clusters have  cored density 
profiles with core radii $r_c \sim 300 \kpc$. Consequently, 
the associated reduced tangential shear lensing profiles  consistently tend to zero 
at angles $\theta \simlt 40^{''}$.
This result is inconsistent with what is deduced from the measured profiles.
These profiles exhibit  a diverging behavior when $\theta \rightarrow 0$, 
as predicted by an  NFW mass model.
We argue that such  contradictions 
cannot be easily reconciled within the  DM models
presented so far as an alternative to the collisionless paradigm.
However, we  suggest   that this tension  can be used as a 
unique test bed to probe new DM physics.
}

\notoc

\begin{document}

\maketitle

\section{Introduction}
\label{sec:intro}

Major mergers between massive galaxy clusters can be considered  as the most 
energetic events since the Big Bang \citep[see][for a review ]{Molnar16}.
The  collisional energy  in these mergers ($ \sim 10^{63} - 10^{64} \ergs$) 
can therefore be used profitably to study the collisional properties of dark matter (DM).

A significant effect expected to arise as a consequence of a
cluster collision is the spatial separation of the collisionless
components (galaxies and DM) from the dissipative intracluster medium (ICM).
Furthermore, an additional offset between the galaxy component  and DM 
is expected if DM is self-interacting (SIDM). 
The amount of this offset will clearly depend on the  DM cross-section, and 
useful constraints on SIDM models \citep{Tulin18} can then be derived 
from measurements of spatial offsets in merging clusters.

The number of observations of  major cluster mergers has steadily increased
over the years \citep{Molnar16}, with the most famous example being 
the  Bullet Cluster  1E0657-56   \citep{Mark02,Clowe06}. 
Lensing measurements revealed  a distinct  spatial separation 
between the collisionless DM component and the X-ray emitting ICM, 
proving  for the first time the existence of DM.

 N-body/hydrodynamical simulations  have proven to be a powerful tool  for
 studying binary cluster mergers.  In this framework the two merging halos are 
 initially separated and in equilibrium,  and their collision evolution is 
 followed in time to model a specific merging event \citep{Molnar16}.
Some examples of such simulations  taken from the literature 
concern the Bullet Cluster \citep{Spr07,Mas08,La14}, the cluster ACT-CL 
J0102-4915 \citep{Donnert14,Molnar15,Zh15,Zh18},
and   the Sausage Cluster \citep{Donnert17,Mol17}.

A very interesting example of such extreme collisions is the  cluster 
ACT-CL J0102-4915  (`El Gordo') at $z=0.870$. This merging cluster was 
originally discovered by the Atacama Cosmology Telescope (ACT) survey
through its Sunyaev-Zel'dovich (SZ)  effect.

The total mass is of about  $\sim 2 \cdot 10^{15} \msun$, 
with a galaxy velocity dispersion $\sigma_{gal} \sim 1,300 \kms$ and an 
integrated temperature $T_\mathrm{X}\simeq 15 $ keV  \citep{Men12}.
The mass estimates are consistent with independent weak lensing 
\citep[WL;][]{Jee14} and strong lensing (SL) studies \citep{Zi13}.
These mass measurements demonstrate that  El Gordo  is the most massive 
cluster at $z>0.6$; an important consequence is that 
the existence of such a massive cluster  at this high  redshift
is difficult to  reconcile within  the standard \lcdm~model 
 \citep{Men12,Jee14,As21,As23}.

The merging system is characterized by two subclusters \citep{Jee14}:
due to their positions these are termed the northwestern (NW) and southeastern (SE), 
respectively \citep[see, for example, Figure 1 of ][]{Ng15}. 
The two subclusters exhibit a projected separation  of  
  $ d \sim 700 \kpc$, with a  mass ratio of  $\sim 2:1$   
  and an infall relative velocity in the range 
  $\sim 1,500\kms$ to $\sim 2,500 \kms$ \citep{Men12}.

  The development of large spatial offsets between the  mass and X-ray 
  peaks of the merging clusters is one of the most interesting effect that
  is expected to take place in high-velocity mergers, 
  such offsets are similarly predicted  between the   SZ and X-ray centroids. 
  \citep{Molnar12}. 

The peak location   of the different mass components  in the El Gordo 
cluster \citep{Men12,Jee14, Zi13,Die20,Kim21} presents several significant 
features.
The most interesting feature is the spatial location of the X-ray peak
of the SE cluster. As expected, it is spatially displaced from  the
DM peak. However, contrary to dissipative arguments
and observations in the Bullet Cluster, the X-ray peak  actually precedes the 
DM peak. Specifically, the X-ray emission peak 
 is farther from the system center-of-mass than 
the corresponding DM mass centroid. 
Additionally,  the  brightest cluster galaxy (BCG)  is also  spatially offset from 
the mass centroid.  It is worth noting that the presence of galaxy-DM offsets in 
major mergers is a specific prediction of SIDM models \citep{Kim17}.
 
X-ray observations reveal a well-defined X-ray morphology, 
with a strong  X-ray emission  peak in the SE region and an 
elongated twin-tailed structure extending beyond the peak. 
 The total X-ray luminosity is $L_X\sim 2\cdot 10^{45} \ergs$ in the 
$0.5-2$ keV band \citep{Men12},  with the NW region  
having a much weaker  X-ray emission.
The  presence of  distinct X-ray morphological features 
suggests that  the  merging  is  approximately taking place 
in  the plane of the sky \citep{Men12}.

To summarize, a coherent  scenario consistent with the above observational findings 
suggests that El Gordo is an high redshift cluster which is undergoing a  major merger.
The simplest model to describe the merger is one in which the two subclusters
collided at high velocity ($ \simgt 2,000 \kms$) and are now in a
post-pericenter phase, moving away from each other. This is the so-called outgoing 
scenario \citep{Men12,Jee14}.

Accordingly, the two-tail cometary structure  and the  wake seen in the 
X-ray images are induced by the motion of the dense, cool gas core of the secondary as 
it moves through the ICM of the primary from  NW to  SE. 
There is not an X-ray peak for the NW cluster because the primary's original gas core
was destroyed during the collision with the compact SE gas  cool core.  
Overall, these findings support the view  of the El Gordo cluster  as 
 an extreme merging event exhibiting  very interesting properties.

 Several authors \citep{Donnert14,Molnar15,Zh15,Zh18,V24} have  carried out  
  N-body/hydrodynamical merging simulations, with the purpose of reproducing the 
  various observational features of this
 merging cluster.
 A series of merging simulations were performed in a collisionless CDM scenario 
 by \citep{Zh15}.
According to the authors, the merging model (``model B'')  that best matches 
observations has a  total mass $\sim 3 \cdot 10^{15} \msun$ and a high mass 
ratio ($\sim 3.6$).  The initial conditions are those of an off-axis merger, 
with  an initial relative velocity between the two subclusters  of $\sim 2,500 \kms$
and impact parameter $\sim 800 \kpc$.

However, the main shortcoming of this merger model
is that most of the X-ray observations are well reproduced 
 for a  primary's cluster mass of about $\sim 2.5 \cdot 10^{15} \msun$.
 This value for the mass of the primary is in tension with more recent 
 lensing estimates, based independently on both SL \citep{Die20} and WL 
 studies \citep{Kim21}.  Both the works predict   significantly  lower  cluster mass 
 values  than ($\sim 30- 60\%$) previously estimated \citep{Men12,Zi13,Jee14}.

Therefore, it is interesting to verify whether this range of masses for the El Gordo 
cluster  is consistent with its observed X-ray morphology.
This has been  investigated  in a recent paper \citep{V24}, 
in which we have presented an ensemble of N-body/hydrodynamical simulations 
of the galaxy cluster El Gordo that  include the recently revised  cluster masses.
Here, we review our recent findings obtained from this series of simulations
 \citep{V24}, in particular from the SIDM merging models.
We also present new results obtained by extracting 
reduced tangential shear profiles from the DM halos of the post-collision clusters.

 The structure of the  paper is as follows. 
 We outline the simulation setup  in Section \ref{sec:sims}:
  the construction of the merging initial conditions is briefly described in 
  Section \ref{subsec:icsetp}, and
  the particle model we use to implement DM self-interactions in the simulations
  is introduced in Section \ref{subsec:icsidm},   Section
  \ref{subsec:imag}  describes the procedure used to construct mock X-ray maps 
   and in Section \ref{subsec:mrgmodels} is discussed the choice of the optimal 
   merger model.

The results are presented in Section \ref{sec:results}, with   
 Section \ref{sec:sidmb} presenting 
results from merger simulations   performed in an  SIDM scenario.
Section \ref{sec:glens} discusses  the consistency of the weak 
lensing profiles extracted from the DM halos of the SIDM merging simulations 
against measured profiles.
Finally, Section \ref{sec:discuss} summarizes our main conclusions.
Throughout this work  we use a concordance \lcdm~ cosmology,  with
$\om=0.3$, $\oml=0.7,$ and Hubble constant 
$H_0=70\equiv 100h$\,km\,s$^{-1}$\,Mpc$^{-1}$.

\section{Method}
\label{sec:sims}

We refer to \citep{V24} for a more detailed description of 
the initial condition setup. Our binary merging simulations were performed using 
 an improved SPH numerical scheme for the hydro part,  
coupled with a standard treecode to solve the  gravity problem.
  The Lagrangian SPH code employs an entropy conserving formulation, while 
in the momentum equations SPH gradients  are estimated 
using a tensor approach.
See, in particular,  \citep{V16} for an in-depth discussion about its 
hydrodynamical performances.

\subsection{Initial conditions }
\label{subsec:icsetp}

The masses of the colliding clusters are defined according to 
 $M_{200}$, which correspond to the mass such that within   the radius
 $r_{200}$ the average density is $ 200$ times the cosmological critical density 
 $\rho_c(z)$:

 \begin{equation}
M_{200}=\frac{4 \pi}{3} 200  \rho_c(z) r_{200}^3~,
 \label{mcl.eq}
 \end{equation}
where  $z=0.87$ is the redshift of the El Gordo cluster. 

We denote as  $M_1$ ($M_2$),  the mass of the  primary 
 (secondary)  cluster, with $q=M_1/M_2 \geq 1 $ being the mass ratio.    
To set up the initial conditions of our merging simulations we create 
a particle realization of two individual halos at equilibrium: the 
mass components of each halo consists of DM, gas and eventually a stellar
component.

\subsubsection{ Halo density profiles   }
\label{subsec:icdm}
The DM halo density profiles  are modeled  according to  an  NFW 
profile 
\begin{equation}
\begin{array}{llll}
\rho_{DM}(r)&=&\dfrac{\rho_s}{r/r_s(1+r/r_s)^2}\, ,  &   0\leq r\leq r_{200} \, , \\
 \end{array}
 \label{rhodmins.eq}
 \end{equation}
where  the scale radius $r_s$ is related to $r_{200}$  by $ r_s=r_{200}/c_{200}$,
and $c_{200}$ is the concentration parameter  given 
by the $c-M$ relation of \citep{Du08}.
For the DM profiles outside of $r_{200}$  we implement an 
 an exponential cutoff up to a  $r_{max}=2 r_{200}$ with a scale length 
 $r_{decay}=0.2 r_{200}$.

 A numerical realization of the DM density profile is determined by 
 sampling the  cumulative DM mass profiles with a uniform random number in the 
 interval $[0,1]$ and solving for the radius $r$.
Similarly, for a particle at position $r$ the 
particle speed $v$ is obtained according to a standard acceptance-rejection 
method by numerically evaluating the DM distribution function $f_{DM}(\mathcal{E})$
over a range of energies.
Finally,  the directions of the particle position and velocity vectors are chosen 
isotropically.

The gas distribution is initialized under the assumption  of hydrostatic 
equilibrium within the DM halo. We choose to model the halo gas densities
  according to  the Burkert profile \citep{Bu95}:
\begin{equation}
\begin{array}{llll}
\rho_{gas}(r)&=&\dfrac{\rho_0}{(1+r/r_c)\left[1+(r/r_c)^2\right]}\, ,  &   0\leq r\leq r_{200} \, , \\
 \end{array}
 \label{rhogins.eq}
 \end{equation}
where $r_c$ is the gas core radius and $\rho_0$ the central gas density. 
Additionally, we  also considered  for the gas density profile of the primary
 a non-isothermal $\beta$-model \citep{Donnert14}:
\begin{equation}
\rho_{gas}(r)= \rho_{0}  \left(1+\frac{r^2}{r_c^2}\right)^{-\frac{3}{2} \beta}~.
 \label{rhogbeta.eq}
 \end{equation}

 For a given cluster and a specific profile, the central  density $\rho_0$  is 
 then found numerically by solving for the  gas mass fraction $f_g$ at 
 $r=r_{200}$.
 We then solve the equation of hydrostatic equilibrium to determine the gas 
 temperature at radius $r$, where we assume for the gas an adiabatic index  
 of $\gamma=5/3$.

For the merging simulations that are  supposed to mimic the presence of BCGs
 we initially incorporate in the halos a  star matter component.
The density profile of the stellar component is 
analytically approximated as in  \citep{Mer06}.
The BCG masses $M_{\star,BCG}$ are derived according to  \citep{Kr18}, and
for the range of halo masses under consideration 
$M_{\star,BCG}\sim 2.3 \cdot 10^{12}\msun$.
Positions and velocities of the star particles are determined 
according to the same procedure adopted for DM particles.

The masses of DM and gas particles  are assigned as in 
\citep{VS21}. For example, a typical simulation with an 
halo mass of  $M_{200}\sim  10^{15}\msun$ has 
  $N_{DM} \simeq 3.4 \times10^5 $   DM particles   
and  $N_g \simeq 1.7 \times10^5 $  gas particles  for 
an halo gas mass of $M_{g}\sim  10^{14}\msun$.

\subsubsection{Initial merger kinematics}
\label{subsec:ickin}

Our merging runs  are performed in the $ \{x,y\}$ plane 
of the simulations, with  the center of mass of the two halos being initially 
separated by a distance $d_{ini}=2(r^1_{200}+r^2_{200})$. 
The halos have initial relative  velocity $V$ and 
impact parameter $P$, the center of mass of the two clusters being 
centered at the origin.
The merger dynamical evolution is then  fully determined by the merging 
parameters $ \{M_1, ~q,~ P, ~V\}$.

\subsection{Numerical implementation of self-interacting dark matter}
\label{subsec:icsidm}

Several approaches have been proposed to implement DM self-interactions 
in N-body simulations.  
In our merging runs \citep{V24}  we considered the  simplest  case 
of isotropic and elastic scattering between DM particles; 
we  further simplified the scattering model
by assuming a constant, velocity-independent DM cross-section $\sigma_{DM}$. 
The local scattering probability is determined as in \citep{Vog12}, 
and to be evaluated requires for each DM particle $i$ 
the definition of  a local DM density $\rho_{DM}({\bf r_i}) $:

 \begin{equation}
 \rho_{DM}({\bf r_i}) =\sum_j m^{DM}_j W(|{\bf r}_{ij}|,h^{DM}_i)~,
    \label{rhodm.eq}
 \end{equation}
where $m^{DM}_i$ is the mass of the DM  particle $i$,
 $ W(|{\bf r}_{ij}|,h^{DM}_i) $ is the $M_4$ kernel with compact support, 
 $h_i^{DM}$   is the  DM smoothing length and 
  the summation is over  $N_{nn}=32 \pm 3$ DM neighboring particles. 
    
According to  \citep{Vog12}, within the simulation timestep  $\Delta t_i$ 
the local scattering probability of a  DM particle $i$   with a neighboring 
DM particle $j$ is

\begin{equation}
P_{ij}= m^{DM}_i W(r_{ij}, h^{DM}_i) \frac{\sigma_{DM}}{m_X} v_{ij} \Delta t_i ~,
 \label{pijdm.eq}
 \end{equation}where $v_{ij}=|\bf {v_i}-\bf {v_j}| $ is the relative 
 velocity between particles $i$ and $j$ and ${m_X}$ is the  physical mass 
 of the DM particle.

At each step the total scattering probability of the $i$ particle  
  is $P_i= \sum_j P_{ij}/2$,
where the factor 2 accounts for the other member of the scattering  pair.
  A collision  between particle $i$ with one of its neighbors $j$
  will then take place whenever 
   $P_i \leq x$, where x is a uniform random number in the range $[0-1]$.
When  this condition is satisfied, the post-scattering velocities of the 
DM  pair are

\begin{equation}
\left\{
\begin{array}{lll}
	{\bf u_i} &=&{\bf V}  + (v_{ij}/2) {\bf e}   \\
	{\bf u_j} &=&{\bf V}  - (v_{ij}/2) {\bf e}  ~,
 \end{array}
\right .
 \label{vdmscatt.eq}
 \end{equation}
 where ${\bf V}= ({\bf v_i}+{\bf v_j})/2$ is the center-of-mass velocity,
and $\bf e $ is a unit vector oriented  in a randomly chosen direction.

\subsection{Simulated observations}
\label{subsec:imag}

For any given epoch and viewing direction we extract from the simulations
two-dimensional maps of surface mass density, X-ray surface brightness,  
and SZ amplitude. The maps are evaluated in the observer frame by 
applying two rotation matrices to the simulation frame  \citep{Zh15,V24}.
In particular, we set the angle between the merging axis and the plane of the sky 
($ i= 30 ^{\circ}$) as in  model B  of \citep{Zh15}.

Specifically, the surface mass density  is defined as 

 \begin{equation}
 \Sigma_m(x,y)= \int_{los}  
\left[\rho_{gas}({\bf x})+\rho_{DM}({\bf x})\right] d z~,
\label{smass.eq}
 \end{equation}
 where $\rho_{gas}({\bf x})$ and $\rho_{DM}({\bf x})$ are the gas and DM
 densities at the position $\bf x$, respectively.

Mock X-ray maps are extracted from the simulations
 following \citep{Molnar15}.  To obtain the X-ray surface brightness 
the X-ray emissivity  $\varepsilon(\rho_g,T_{g},Z, \nu )$    
is integrated  along the line of sight and over the energy range
  $[0.5-2]$ keV:

 \begin{equation}
 \begin{split}
         \Sigma_{X}(x,y)  =  &  \frac{1}{4 \pi (1+z)^4}  \int_{los} \,dz  \\
  &        \int \, \varepsilon(\rho_g,T_{g},Z, \nu ) A_{eff}(\nu) \, d \nu  ~,
\end{split}
\label{sbrx.eq}
 \end{equation}
here  $T_g$ is the gas temperature,  $\nu$ the frequency, $Z$ the 
metal abundance of the gas, and $A_{eff}(\nu)$ the effective area of the 
telescope. 
We set for our  mock X-ray maps (\ref{sbrx.eq}) 
 the exposure time to $t_{exp}=60ks$ \citep{Zh15}, 
they are then  expressed in counts arc sec$^{-2}$.

 The  SZ surface brightness  at the frequency $\nu$  is calculated 
 including relativistic corrections \citep{Itoh98}:

 \begin{equation}
 \begin{split}
 \Sigma_{SZ}(x,y)  = &  \frac{\sigma_T k_B}{m_e c^2} \int_{los}  n_e T_{g}  \\
         &       \left[ g(\nu)+ \Sigma_{k=1}^{k=4} Y_k \Theta^k \right]  dz~,
 \end{split}
\label{sbsz.eq}
 \end{equation}

where $\sigma_T$ is the  Thomson cross section,
$m_e$ the electron mass,  $c$ the speed of light,
 $n_e$ the electron number density, $k_B$ the Boltzmann constant, respectively. The function $g(\nu)= \coth(x_{\nu}/2)-4$   is the nonrelativistic 
 frequency function, where $x_{\nu}= h_P \nu /(k_B T_{cmb})$ and $T_{cmb}$ 
 is the  cosmic microwave background temperature. 
The coefficients $Y_n$  are the relativistic corrections as given by 
\citep{Itoh98}, and $\Theta \equiv k_B T_g /m_e c^2$.  
 We smooth the SZ maps with a Gaussian kernel with width $\sigma_{SZ}=270$ kpc 
($\sim0.55^{\prime}$ at $z=0.87$) and set $\nu=150 $ GHz \citep{Zh15}.

The projected maps are  evaluated  on a
 2D mesh of $N_g^2=512^2$ grid points.
The centroid positions of the various maps 
are located by applying  a shrinking circle method to the simulation particles. 

\begin{table*}
\caption{IDs and initial merger parameters of the SIDM merging simulation 
of Figure  \ref{fig:planeSXc}.$^{a}$}
\label{clsdm.tab}%
\centering
\scalebox{0.9}{   
\begin{tabular}{cccccccccc}
\hline
Model & $M^{(1)}_{\star} [\msun]$  & $M^{(2)}_{\star} [\msun] $ 
& $N^{(1)}_{\star} $ & $\varepsilon_{\star} [\kpc] $ & $\rcp [\kpc]$ &
$ \zeta$ & $ \sigma_{DM}/m_X [\sxu]$  & $(f_{g1},f_{g2})$ \\
\hline
XDBf\_{sb} &  $ 2.2 \times 10^{12}$ & $ 1.6 \times 10^{12} $ & $16,785$ &
  9.5 & 290 & 2.4  &  4 &  (0.12,0.14) \\
\end{tabular}
}
\begin{flushleft}
{\it Notes.} {$^a$ Columns from left to right:
 ID of the merging model, stellar mass of the BCG of the primary,  the same mass 
but for the secondary, number of star particles for the primary, 
  gravitational softening  length  of the star particles,
  gas core radius  of the primary, dimensionless parameter $\zeta=r_s/r_c$,   
SIDM cross-section per unit mass, primary and secondary cluster gas mass 
fractions $f_g$ at  $r_{200}$. 
The  collision  parameters of the SIDM merger model 
are those of model Bf in Table 1 of \citep{V24}: 
$\{ M^{(1)}_{200},~q,~V,~P \} =   
\{  1.6 \cdot 10^{15} \msun, ~2.32, ~2,500 \kms, ~ 600 \kpc \}  $
}.
\end{flushleft}
\end{table*}

\subsection{Merger model}
\label{subsec:mrgmodels}
We present results from the SIDM merger model of \citep{V24}
which showed the most interesting observational properties.
For this off-axis merger model the masses of the primary and secondary are 
chosen in accordance with recent mass estimates \citep{Die20,Kim21} and
set to $M^{(1)}_{200}  = 1.6 \cdot 10^{15} \msun $ and
 $M^{(2)}_{200} =  6.9 \cdot 10^{14} \msun $, respectively.
 The initial merger configuration is completed by choosing 
$V=2,500 \kms$ for the initial relative velocity and $P=600 \kpc$ 
as   impact parameter. These merging parameters
 $\{ M^{(1)}_{200}, ~q,~P, ~V \} = 
 \{ 1.6 \cdot 10^{15} \msun , ~2.32,~600 \kpc, ~2,500 \kms \}$ are those
 of model Bf in Table 1 of \citep{V24}. 
The halo concentration parameters  are 
 $\concI=2.5$ and $\concII=2.682$ for the primary and 
 secondary halos, respectively.

 We adopt the $\beta$-model (\ref{rhogbeta.eq})  to describe the initial radial gas 
 density profile of the primary, with   $\beta=2/3$ and gas scale radius
 $r_c=290 \kpc$.  The central density $\rho_0$  is determined by setting 
   the primary's  cluster gas fraction to $f_{g1}=0.12$. 
   The initial gas density of the secondary is instead modeled according to  
 the Burkert profile (\ref{rhogins.eq}),  with  gas core radius  set 
 to $r_c=r_s/3 \sim 164 \kpc$ and the  gas fraction  to $f_{g2}=0.14$. 

Finally, to mimic the presence of a BCG a stellar component is initially added to the  
mass distribution of each individual halo.
We performed a particle realization of the star density profiles
using the procedures described in Section \ref{subsec:icsetp}.
Accordingly, we obtain $M^{(1)}_{\star} = 2.2 \cdot 10^{12} \msun $ 
  and $M^{(2)}_{\star} = 1.6 \cdot 10^{12} \msun $ for the stellar masses of 
  the primary and secondary cluster,  respectively.

For this set of initial collision parameters we performed  an SIDM merger 
simulation  with ${\sigma_{DM}}/{m_X} = 4 \sxu$.
Table \ref{clsdm.tab} summarizes some merger parameters, we use the same 
notation of \citep{V24} and label the simulation as XDBf\_{sb}.

\begin{figure*}[!ht]
\centering
\includegraphics[width=0.95\textwidth]{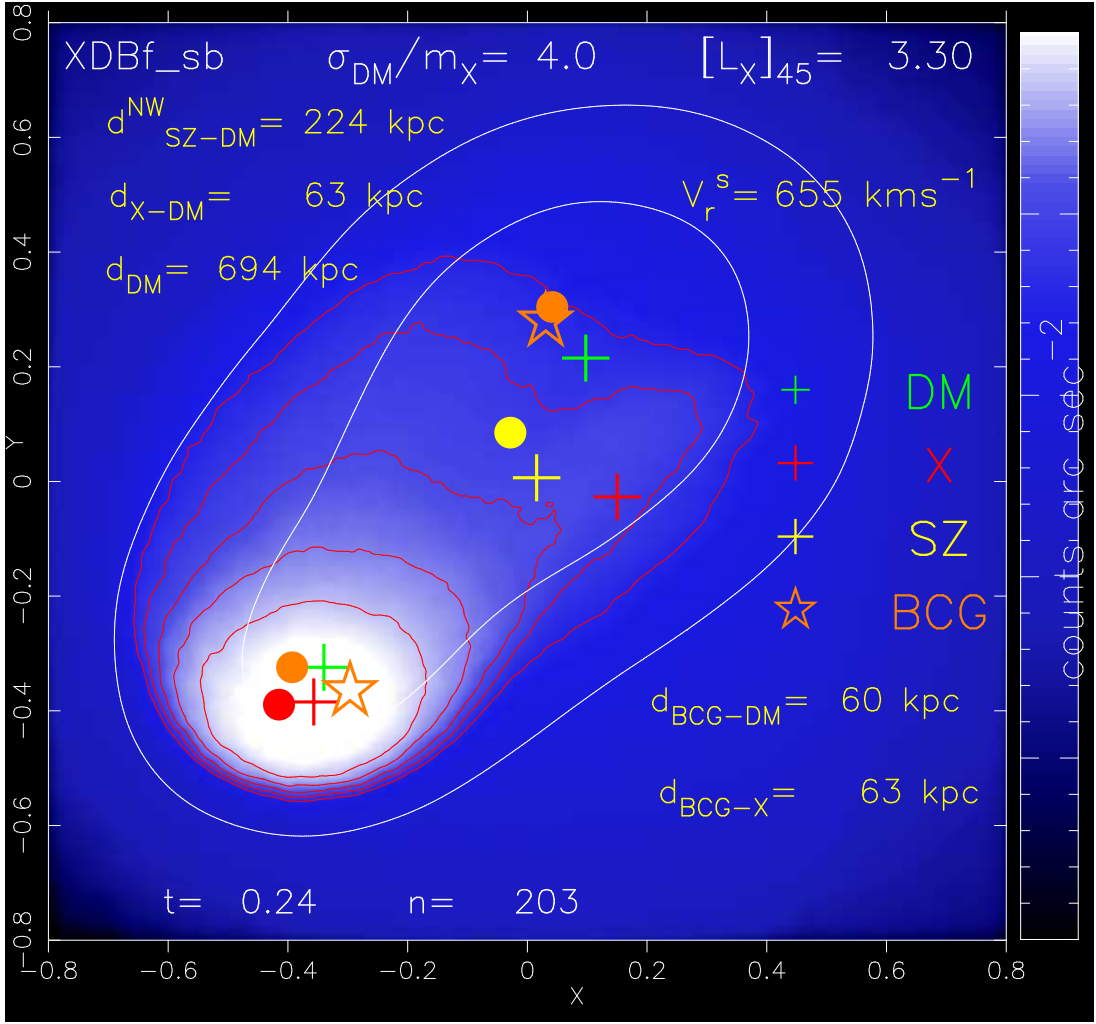}
\caption{X-ray image extracted from the SIDM merging simulation XDBf\_sb 
 at the observer epoch, $t=0.24 $ Gyr after  the pericenter passage.
The box size is $1.6  \mpc$ and  the initial collision parameters are those of
model Bf in Table 1 of \citep{V24} ( see text).
The log-spaced contour levels of the
projected X-ray surface brightness (red) and mass density (white) are shown 
overlaid.
From the inside to outside, the contour levels of the X-ray surface brightness
and of the surface mass density are: $(6.6,4.4,2.9,1.9,1.2)\cdot 10^{-1} \ctsn $
and $(5.6, 3.1,1.8) \cdot 10^{-1} \dms$.
The crosses indicate the projected spatial locations of the mass
(green),  X-ray surface brightness (red) and SZ centroid (yellow).
The  open orange stars mark the  projected spatial location
 of the mass centroids of the star particles representing the BCGs.
 The X-ray luminosity $L_X$ in the $0.5-2$ keV band  is given in units of
$10^{45} \ergs$.   
The distance $d_{\XDM}$ indicates the value in kpc of 
the projected  distance between the X-ray emission peak and the DM mass 
centroid, $d_{\BX}$ that between the mass centroid of the BCG galaxy and 
the X-ray emission peak, and finally $d_{\BDM}$ is the distance between 
the BCG and DM mass centroids. All of the centroids 
refer to the SE cluster, with 
the exception of $d^{NW}_{\SDM}$, which is the projected distance between 
the SZ peak and the DM mass centroid of the NW cluster.
The value of $V^s_r$ refers to the line-of-sight relative 
mean radial velocity between the two BCGs. 
The filled circles indicate the peak locations from several observations, as
taken from Figure 6 of  \citep{Kim21}. Their spatial positions have been
normalized to the relative distance from the mass centroids. 
The color coding of the circles is the same of the associated crosses, 
which indicate the projected positions of the corresponding centroids
as extracted from the simulation. 
}
\label{fig:planeSXc}
\end{figure*}

\section{Results}
\label{sec:results}
Section \ref{sec:sidmb}  provides a  review  of the key findings 
of the SIDM merging model XDBf\_{sb} presented  in Section 3.3 of \citep{V24}.
 Section \ref{sec:glens} discusses the consistency of the weak 
lensing profiles extracted from the DM halos of the SIDM merging simulation 
with  measured profiles  as obtained from available lensing measurements.

\subsection{El Gordo merger with SIDM}
\label{sec:sidmb}

 Figure \ref{fig:planeSXc} shows the mock X-ray map extracted from  the
 SIDM merger simulation.  
One  significant conclusion that can be drawn from the 
 map of Figure \ref{fig:planeSXc} is  the behavior 
 of the X-ray gas morphology in an SIDM merger.
The contour levels of the projected mass density are  much rounder 
 than those  extracted from the corresponding standard CDM merger 
 model \citep{V24}.
 This indicates shallower DM potential wells, which in turn lead  to a
 reduced resiliency  of the post-pericenter gas structures, 
which can now more easily escape from the potential wells of the 
original halos. 

This is a specific signature of SIDM: due to  DM interactions the expected 
exchange of energy during the collision between the two clusters
will result in  shallower 
 DM halo potential wells \citep{Kim17,ZuH19}.
Consequently, the X-ray emission in the outer regions behind the 
secondary is significantly reduced compared to  the measured emission 
 of the analogous collisionless CDM merger model 
 \citep[See Figure 10 of ][]{V24}.

 We also show  in Figure \ref{fig:planeSXc} the positions of 
 the different centroids, as extracted from the simulation. 
The crosses indicate the projected spatial locations of the DM 
(green) and X-ray emission peak (red)  centroids. 
The  open orange stars refer to the  projected spatial location of the mass 
centroids of the star particles representing the BCGs, the yellow cross shows 
the position of the SZ centroid.

We also report the measured positions of the different centroids.
These are extracted from Figure  6 of \citep{Kim21}, and for each 
cluster their positions are relative to the location of the corresponding
mass peak.
These relative positions are indicated in Figure \ref{fig:planeSXc} 
with filled circles, the color coding being the same of the corresponding
centroids extracted from the simulation.
For the SE cluster  are indicated the distance of the 
 DM to the X-ray peak ($d_{\XDM}$), the BCG to the X-ray peak ($d_{\BX}$) 
 and that of the BCG to the DM centroid ($d_{\BDM}$), 
The distance of the SZ to the DM centroid ($d^{NW}_{\SDM}$)
 refers to the  NW cluster.

 The magnitude of the different offsets can be used to set constraints on 
  ${\sigma_{DM}}/{m_X} $, a critical issue being  the observational 
  uncertainties in the measured positions of the various centroids.
According to \citep{Kim21}, the null hypothesis of zero size offsets
can be   excluded with high significance.

The most significant offset is the position of the X-ray peak of the 
SE cluster, which is located further away from the system center-of-mass than 
the corresponding DM mass centroid. 
This is clearly in tension  with what is expected 
in a  collisionless CDM scenario,  but it can be  natural explained   by a 
SIDM merger model.
  Figure \ref{fig:planeSXc} shows that   $d_{\XDM} \sim60 \kpc $
 which, within the  observational scatter (see below), can be considered in 
 accord with the measured offset $d^{SE}_{X-DM} \sim 100 \kpc $.
 Similarly, Figure \ref{fig:planeSXc}  shows an offset  of the SZ peak 
from the NW DM  centroid of about $d^{NW}_{SZ-DM} \sim 230 \kpc $. This value
is in better agreement with data and significantly lower than 
 the values found in  standard CDM merging  runs \citep{Zh15,V24}. 

As can be seen from Figure \ref{fig:planeSXc}, the BCG  of  the SE cluster
exhibits an offset of about  $d_{\BDM} \sim 60 \kpc $  from the DM 
centroid, which is in the same range of the observed one ($\sim 60 \kpc$).
These offset are expected in an SIDM  scenario, during the cluster collision the DM 
halos will  experience an exchange of energy and in turn a deceleration, thus 
leading to the formation of positive BCG-DM offsets.

 We therefore conclude that  these findings are among the  most interesting 
 results of our study,  and strongly support an SIDM merger model for the 
 El Gordo cluster.

 Finally, after the pericenter passage the gravitational pull of the DM halos
  will begin to  reduce the BCG bulk velocities. As a result, 
the relative mean radial velocity  $V^s_r$ along the line of sight 
between the  two  BCGs is now of the order of $V_r^s\sim650 \kms$ 
(see Figure \ref{fig:planeSXc}), much lower than in the collisionless 
CDM cases ($V_r^s\sim1,000 \kms$, see Figure 7 of \citep{V24}) and 
 in  better agreement with the measured value of
  $V_r^s=598 \pm 96 \kms$ \citep{Men12}. 
This is clearly another positive feature of the  SIDM  merger model 
presented here.

\subsection{DM halo density profiles and averaged radial lensing profiles}
\label{sec:glens}

\begin{figure*}
\centering
\includegraphics[width=17.2cm,height=8.2cm]{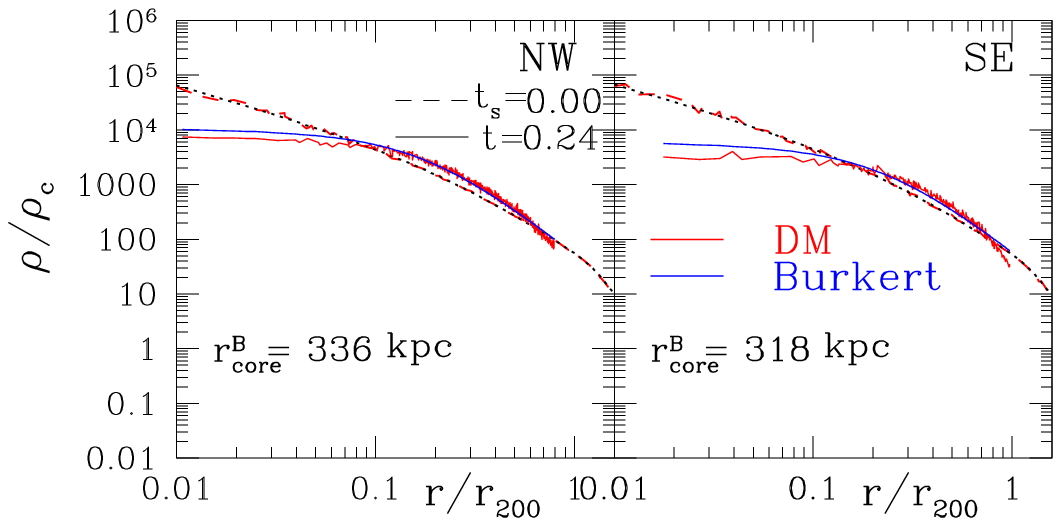}
\caption{Measured radial  density profiles  of the two DM halos 
of  the merger model XDBf\_sb  of Figure \ref{fig:planeSXc}.
The left (right) panel is for the NW (SE) cluster. 
Solid red lines refer to the present epoch, which is at $t=0.24$ Gyr after the 
pericenter passage, at this time  the projected separation  
between the mass centroids of the two components is approximately  
$ d_{DM} \sim 700 \kpc$. The dashed red lines correspond to the 
simulation time $t_s=0$, at the start of the simulation. 
An NFW density profile is used to fit the DM density profile of each cluster 
  at $t_s=0$ (black dot line), while in order to fit the cored DM  profile at 
$t=0.26$ Gyr  we adopted a Burkert profile (solid blue line).    
In each panel is reported the value of the corresponding core 
radius $r^B_{200}$, the related   statistical error being   negligible.
\label{fig:profDMh}
}
\end{figure*}

One of the main effects of collisional DM is the development of cored DM
halo density profiles. Moreover, the dependence of the scattering
  probability (\ref{pijdm.eq}) on the relative velocity between DM particles
  implies that this effect will be  further enhanced in SIDM cluster mergers.
  This is because the  relative collision velocity between the two clusters
 being much higher ($v_{rel}\sim 4,000 \kms$ at  the pericenter) 
 than the velocities expected from the internal motions of an isolated halo.

Figure \ref{fig:profDMh} shows the radial density profiles of the two cluster DM halos
for our SIDM merger model XDB\_sb.
These are plotted in the left (right) panel for the NW (SE) cluster.
Solid  lines refer to the observer epoch ($t=0.24$ Gyr) and 
dashed lines  to the start of the simulation.
It can be seen that, in accordance with SIDM predictions, 
at the observer epoch the two DM halos exhibit  flattened density profiles
 in their inner regions,  with core radii of approximately  $ \sim 300 \kpc$.

These  post collision cored DM density  profiles are better modeled  
 using a theoretical profile that includes  a core radius as one of  its profile 
 parameters.
We found analytically convenient (see later) to use the Burkert profile
 (\ref{rhogins.eq}) to model the  DM density  profiles  shown in 
Figure \ref{fig:profDMh}.  We then fitted these profiles according to
 the analytic model (\ref{rhogins.eq}),  with  $r_c$ 
being  now the DM core radius and $\rho_0$ the central DM density. 
The resulting best-fit profiles  are  depicted as solid blue lines in 
Figure \ref{fig:profDMh} and,  as can be seen from the Figure,  
the chosen modeling turns out to provides  a better  fit to the 
 DM halo profiles than the previous NFW model.

We want now to compare the DM halo density profiles, as predicted by our 
merger model, against possible constraints as derived from weak lensing studies 
 of the El Gordo mass distribution \citep{Jee14,Kim21}.
A fundamental quantity to probe the cluster mass distribution in the WL
 regime \citep[see][for a review ]{Um20}  is 
the projected mass distribution. This is obtained at the projected radius $R$ 
from the 3D matter density: 
\begin{equation}
\Sigma(R)= \int_{-\infty}^{\infty}  \rho( \sqrt{R^2+z^2})dz ~.
 \label{Sigma.eq}
 \end{equation}

The averaged  surface mass density within the circle of  radius $R$ is
accordingly defined as 

 \begin{equation}
 \bar {\Sigma} (R)  =  \frac{2}{R^2} \int_0^R  \Sigma(R^{\prime}) 
R^{\prime} d R^{\prime}  ~,
\label{sbrav.eq}
 \end{equation}
and the excess surface mass density is then obtained  as
$\Delta \Sigma(R) \equiv \bar {\Sigma}(R) - \Sigma(R)$.

In the WL regime the  tangential shear $\gamma(R)$ is related to the 
 excess surface mass density \citep{Um20} by 

\begin{equation}
\Delta \Sigma(R) = \gamma(R)  \Sigma_c~,
\label{gamma.eq}
 \end{equation}

where  $\Sigma_c$ is the critical surface mass density:  

\begin{equation}
\Sigma_c \equiv \frac{c^2} { 4 \pi G } \frac{D_s} {D_d D_{ds}}~,
\label{sigmac.eq}
 \end{equation}

and $D_s$ , $D_d$  , and $D_{ds}$ are the angular diameter distances between
the observer and the source, from observer to lens, and from the lens to the
 source, respectively.

\begin{figure*}
\centering
\includegraphics[width=17.2cm,height=15.8cm]{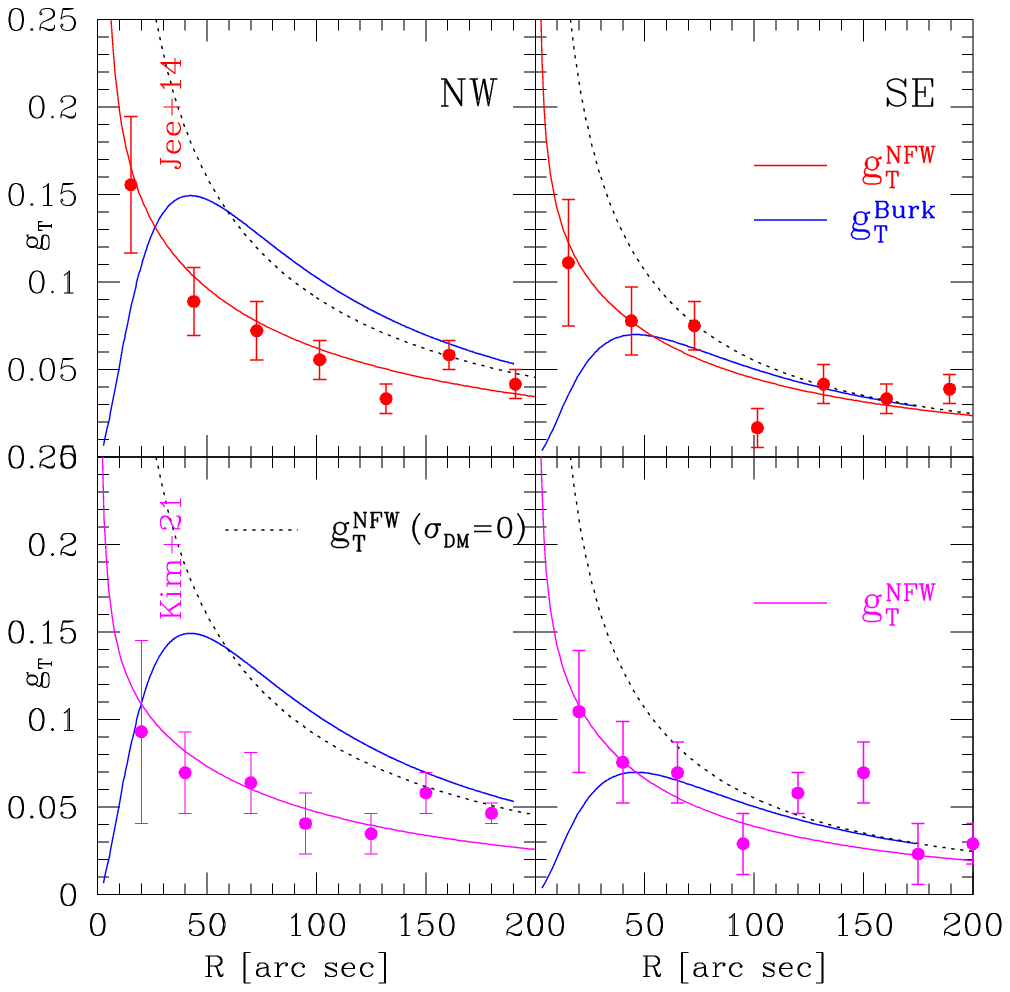}
\caption{Left (right) panels show the reduced tangential  shear profiles
  $g_T(\theta)$  for the  NW (SE) cluster, as measured by some authors. 
Top and bottom panels indicate the data points as extracted from Figure 9 
of \citep{Jee14} and Figure 17 of \citep{Kim21}, respectively.
In each panel the data points are  compared against a
$g^{NFW}_T(\theta)$ profile as obtained by a  NFW mass model. 
For the top panels the $g^{NFW}_T(\theta)$ profiles are constructed 
using the best-fit NFW parameters taken from Table 2 of \citep{Jee14}: 
$ \{ r_{200}^{NW}, r_{200}^{SE} \}= \{1.65,1.38\} \mpc$.
and $\{ \concI, \concII \} = \{ 2.57, 2.65 \}$.
The $g^{NFW}_T(\theta)$  profiles shown in the bottom panels are computed 
according to the NFW parameters reported in  Table 2 of \citep{Kim21}:
$ \{ r_{200}^{NW}, r_{200}^{SE} \}= \{1.5,1.3\} \mpc$ 
and $\{ \concI, \concII \} = \{ 2.54, 3.20 \}$.
Solid blue lines refer to the reduced tangential  shear profiles
$g^{Burk}_T(\theta)$, these have been inferred  from  the best-fit 
Burkert density profiles  used to model the cored DM profiles 
extracted from the SIDM merging simulation XDBf\_sb and 
 shown in Figure \ref{fig:profDMh}.
The black dot lines correspond to the NFW lensing profiles 
$g^{NFW}_T(\theta)[\sigma_{DM}=0]$.
These were  derived from an  NFW density model   used to reproduce 
the final halo  DM density  profiles of a mirror simulation of  model XDBf\_sb. 
The simulation was performed  by adopting the same initial condition setup of the
SIDM merging run  XDBf\_sb, but without allowing DM to be self-interacting
by setting $\sigma_{DM}/{m_X} =0$. The NFW parameters of the density profiles are
$ \{ r_{200}^{NW}, r_{200}^{SE} \}= \{1.84,1.38\} \mpc$
and $\{ \concI, \concII \} = \{ 3.97, 5.0 \}$, respectively.
\label{fig:profGT}
}
\end{figure*}

What is observationally relevant is the azimuthally averaged reduced tangential 
shear profile $g_T(R)$, measured around the center of each cluster. 
For the assumed cosmology  $\sim 10^{''}$  correspond to $ \sim 80 \kpc$ at 
the cluster redshift, and we can express the radial dependency of the 
profiles in angular coordinates:

\begin{equation}
g_T(\theta)=\frac{\gamma(\theta) }{ 1 -\kappa(\theta)}~,
\label{gshear.eq}
 \end{equation}
where $\kappa=\Sigma(\theta)/\Sigma_c$ is the WL convergence and
in the WL approximation $\kappa<<1$.

For the NW and SE cluster, Figure \ref{fig:profGT} shows the measured reduced 
tangential  shear profiles $g_T(\theta)$. The binned data are extracted from 
 Figure 9 of \citep{Jee14} (top panels), and Figure 17 of \citep{Kim21} 
(bottom panels).
The measured profiles are compared against an NFW mass model, and for a given NFW 
density profile the corresponding lensing profile $g^{NFW}_T(\theta)$ can be 
calculated analytically.  We refer to \citep{Wri20,Um20} for a derivation 
of the functional form of $g^{NFW}_T(\theta)$.

For each cluster we show in the corresponding panel the profile 
$g^{NFW}_T(\theta)$. This is  computed according to the best-fit parameters 
of the NFW model applied by the authors to describe the cluster mass distribution
(see the caption of  Figure \ref{fig:profGT}). 
All of the NFW lensing profiles are consistently normalized according 
to $\Sigma_c \simeq 4050  \msun \pc^{-2}$ \citep{Jee14}.

We also compare the measured lensing profiles against  those derived by the 
 DM radial density profiles of the clusters, as predicted by our 
SIDM merger model.  To this end, we use  the Burkert profiles  
previously employed to model the cored DM density  profiles  
shown in Figure \ref{fig:profDMh}. Accordingly,  the surface mass density 
(\ref{Sigma.eq})  is then given by 
\begin{equation}
\Sigma_B(R)= 2 r_c \rho_0 \int_0^{\infty} \frac{ dz^{\prime}} { (1+s)(1+s^2)}
\equiv 2 r_c \rho_0  I(u)~,
 \label{SigmaB.eq}
 \end{equation}
where $s^2=z{^\prime}^2+u^2$, $z{^\prime}=z/r_c$ and $u=R/r_c$. 
Over the range of interest, from $u=0$ to $u\sim0.8 r_{200}/r_c\sim4$, the 
 integral $I(u)$ is well approximated to within a few percent by 

\begin{equation}
I(u)\simeq \frac{\pi}{4} \frac{1}{1+u^2}~.
 \label{intpr.eq}
 \end{equation}

Therefore, $\Sigma_B(R)$ reduces to 

\begin{equation}
\Sigma_B(R)\simeq 2 r_c \rho_0  \frac{\pi}{4} \frac{1}{1+u^2}~,
 \label{SigmaBr.eq}
 \end{equation}
and the excess surface density becomes

 \begin{equation}
 \bar {\Sigma}_{B} (R)   \simeq 2 
 r_c \rho_0 \frac{\pi}{4}   \frac{1}{u^2}  \ln (1+u^2) ~.
\label{sbrx2.eq}
 \end{equation}

For each cluster, the lensing profile $g^{Burk}_T(\theta)$ can now be 
calculated by using  the best-fit parameters of the corresponding 
Burkert density profile shown in Figure \ref{fig:profDMh}.  
The Burkert lensing profiles are consistently rescaled using the same value of the
critical surface mass density $\Sigma_c$ previously employed to normalize 
the NFW lensing profiles. 

For the sake of completeness, we also show for the NW and SE cluster the NFW 
lensing profile $g^{NFW}_T(\theta)[\sigma_{DM}=0]$. 
These profiles were   calculated according  to the best-fit NFW density profiles
 used to model at the present epoch the radial  DM density  profiles
of a merging cluster simulation.
The simulation was performed  by adopting the same initial condition setup of the 
SIDM merging run  XDBf\_sb, but without allowing DM to be self-interacting 
by setting $\sigma_{DM}/{m_X} =0$.
 
From the left panels of Figure \ref{fig:profGT},  it can be seen that for the 
NW cluster the simulated lensing profiles $g^{Burk}_T(\theta)$ and 
$g^{NFW}_T(\theta)[\sigma_{DM}=0]$ are significantly offset from the 
shear profile $g^{NFW}_T(\theta)$ derived from lensing data.
The two profiles diverge from each other in the inner cluster region 
($\theta \simlt 50^{''}$), but at larger angles are systematically higher than 
$g^{NFW}_T(\theta)$, with the difference  being of  about a factor of $1.5-2$  
at $\theta \sim 200^{''}$.

This tension can be  understood  as a consequence of the assumed initial 
masses  $ M_{200} $  for the two colliding clusters. 
For the SIDM merging simulation XDBf\_sb the initial collision parameters 
  are the same   of model Bf in Table 1 of \citep{V24}: 
 $ \{ M^{NW}_{200},  M^{SE}_{200} =\{ 16 ,  6.9  \} \times 10^{14} \msun $ 
for the NW and SE cluster, respectively.
These values can be compared with the corresponding mass estimates 
reported in  Table 2 of \citep{Jee14} and \citep{Kim21}: 
 $ \{ M^{NW}_{200}(J),  M^{SE}_{200}(J) =\{ 13.8 \pm 2.2,  7.8 \pm 2 \} 
\times 10^{14} \msun $  and
 $ \{ M^{NW}_{200}(K), M^{SE}_{200}(K) \} = \{ 9.9^{+2.1}_{-2.2}, 6.5^{+1.9}_{-1.4} \} 
\times 10^{14} \msun $,  respectively.

Moreover, for the SIDM simulation  the present cluster masses at $r=r_{200}$ 
are found to be higher by a factor of $\sim 10-30\%$ compared to their
initial values:  $ \{ M^{NW}_{200},   M^{SE}_{200} \}\sim \{ 20 , 7.9 \} 
\times  10^{14} \msun $.
 This is due to the flattening of the DM inner density profiles
 during the merger, and at the present epoch  this in turn leads   
to an average higher DM density at large cluster radii.
Therefore  the differences shown in  Figure \ref{fig:profGT}
at large angles between the lensing profiles of each cluster 
can be simply understood in terms  of the various masses $ M_{200}$  
used to model the corresponding cluster mass distribution.

From  Figure \ref{fig:profGT}  it can be seen that at small angles 
($\theta \simlt 50^{''}$)  there is a significant discrepancy between 
 the NFW lensing profiles $g^{NFW}_T(\theta)$ and the profiles 
$g^{Burk}_T(\theta)$  extracted from the SIDM simulation. The former are
derived from lensing data and  within the allowed uncertainties 
consistently  exhibit a divergent behavior, with 
$g^{NFW}_T(\theta)\rightarrow \infty $ as $\theta \rightarrow 0$.
This is at variance with the angular dependency  of the Burkert lensing 
profiles, which for $\theta \simlt 50^{''}$  start to decrease 
and  $g^{Burk}_T(\theta)\rightarrow 0 $ as $\theta \rightarrow 0$.
  
This is not surprising, as the lensing profiles $g^{Burk}_T(\theta)$  are
derived from the  Burkert density profile (\ref{rhogins.eq}), 
which has been  specifically employed to model the cored DM density profiles
seen in the SIDM merging simulation.
The inconsistency between $g^{NFW}_T(\theta)$ and  $g^{Burk}_T(\theta)$  
as $\theta \rightarrow 0$ is highly significant for both of the clusters, 
we are then forced to conclude
 that the SIDM merging simulation  presented here  
cannot satisfy the constraints inferred from WL data in the 
 El Gordo cluster  inner regions. 

This is in contrast with our previous  conclusions \citep{V24}, 
according to which statistical uncertainties in the reconstructed mass 
profiles \citep{Kim21}  did not allow to rule out 
the presence of cored DM profiles in the El Gordo cluster.
In the next section we will discuss the implications of these findings 
for SIDM merging models of the El Gordo cluster.

\section{Conclusions}
\label{sec:discuss}

In this talk we reviewed  the main results, that were previously 
presented in \citep{V24}, of a simulation study of the merging cluster El Gordo.
Additionally, we also discussed the consistency of the WL profiles,  
 extracted from the DM halos of the SIDM merging simulation, 
against measured lensing profiles.
A summary of our main findings in a collisionless CDM scenario 
is as follows:

i) The observed  twin-tailed X-ray morphology, as well as other observational 
constraints,  are well  matched  by  off-center fiducial merger models  
(see Table 4 of \citep{V24}) with mass of the primary between 
$\sim 10^{15} \msun$ and $\sim 1.6 \cdot 10^{15} \msun$, collision velocities 
and impact parameters in the range 
$2,000 \kms  \simlt V \simlt 2,500 \kms$  and  
$600 \kpc \simlt P \simlt 800 \kpc$, respectively.

ii) One of the most significant features of the galaxy cluster El Gordo is the 
spatial location of the X-ray emission peak, which is further offset from the 
center-of-mass than the corresponding SE DM centroid. 
A returning scenario  was proposed by \citep{Ng15}  as a possible solution to 
this 
issue, according to the authors the merging cluster is
observed in a post-apocenter phase with the two cluster DM  halos 
now  moving toward each other and the SE X-ray peak  moving in the 
opposite direction.

This scenario has been thoroughly studied in Section 3.2 of \citep{V24}, the
conclusion being that the likelihood  of a returning scenario matching 
 the observational constraints from the X-ray morphology of 
 the merging cluster El Gordo is very low.
This conclusion  follows because hydrodynamical simulations showed 
that the orbital time necessary to the DM halo of the SE cluster to reach the 
apocenter and return  ($ \simgt 1 $ Gyr), turns out to be much higher than 
the lifetime ($ \sim 0.1-0.3 $ Gyr) of the post-collision X-ray 
structures.

iii) Two of the fiducial models of  point i) were re-simulated 
to   mimic  the presence of BCG's ( Section 3.3 of
\citep{V24} ). This was achieved by adding a distribution of star particles
to the initial mass components of each of the two halos.

The results of these simulations showed  that, at the observer epoch, there 
were no  significant  differences between the positions of the BCG centroids 
relative to those  of the  DM halos.
In a collisionless CDM merger model of the El Gordo cluster this leaves 
open  the question of the observed BCG to DM offsets.
In principle, such offsets cannot be ruled out as a consequence of 
 violent cluster collisions \citep{Martel14}, 
 although it remains unclear how the gas structure of the 
SE cluster could survive  a cluster collision sufficiently strong to displace
the BCG from its original position at the center of the DM halo.
 Finally, it is worth noting that in a   returning scenario there is no  
 clear  explanation for the observed BCG to DM separation.

iii) For  the standard CDM merger models of point i)  
another problem are the mean relative velocities along the line of sight 
between the SE and NW  BCG components. These values are  significant  higher 
($V^s_r\sim 1,000 - 1,200 \kms $)  than the measured value of 
$V_r^s\simeq 600 \kms$ \citep{Men12}.

Overall, these findings support the study of SIDM merger models for the 
El Gordo cluster.
 In Section \ref{sec:sidmb}  We presented the SIDM merger model with the most 
interesting properties among the  previously discussed merger cases \citep{V24}.
Our main findings can be summarized as follows:

i) The most important results  emerging  from the SIDM merger model 
 XDBf\_{sb} of the El Gordo cluster is that a  simulation with a  DM cross-section 
 of the order of ${\sigma_{DM}}/{m_X} \sim 4\sxu$  can match 
 the observed spatial separations  between the different 
 peak locations.

 However, in order to draw statistical meaningful conclusions it is
 first necessary to assess the statistical significance of the 
 observed offsets. To this end,  we will now attempt to  estimate  the positional 
 error of the X-ray emission peak of the SE cluster. The   
 corresponding offset   can be clearly considered 
 as the most significant of the merging system.

Because of the squared dependence of the X-ray emission with the gas density,
the peak positional error is expected to be relatively small  and determined 
by the angular resolution ($\sim 0.5^{''}$) of the \textit{Chandra} X-ray image.
The WL uncertainty  in the mass peak position,  $\sigma_{DM}\sim 40 \kpc$, 
is then the biggest source of error in determining  the observed separation
between the X-ray  and the SE mass peaks; 
as a result we can estimate the offset to lie  in the range 
$d^{SE}_{X-DM} \sim 100 \pm 40 \kpc $.
According  to \citep{V24}, this constraint on the measured offset cannot be 
satisfied by SIDM merger simulations of the El Gordo 
cluster with ${\sigma_{DM}}/{m_X} \simlt 2 \sxu$.

ii) Another interesting feature  of this merger model is the value of 
 the  relative   radial velocity between the two BCGs. This is 
 at variance  with the findings of standard CDM mergers   
and  is now of the order of several hundred $\kms$,  no longer in 
 conflict with observations.
As previously outlined  in Section \ref{sec:sidmb},  
during the cluster collision the two DM halos will decelerate because of the 
exchange of energy.  Consequently, after the pericenter passage 
the two BCGs will experience  a gravitational pull as they begin to 
exit the potential well.

The above points emphasize the main benefits  of assuming an SIDM scenario 
to model  the merging cluster El Gordo.
Nonetheless, such a  scenario presents several critical issues 
that remain unresolved  in the proposed merger model.  In the following, we list 
and discuss these critical aspects.

iii) The SIDM merger model  presented in Section \ref{sec:sidmb} 
exhibits  a twin-tailed X-ray morphology   which is less defined 
than that observed, even after the adoption of 
initially higher  gas fractions and of a larger gas scale radius for the primary.
This is because the potential wells of the cluster DM halos are much shallower 
than in the collisionless CDM merger, this in turn implies  a reduced 
resiliency  of the post-pericenter gas structures, 
which can now more easily escape from the potential wells of the original halos.

A possible solution to solve this problem is to perform  merger simulations
where the initial cluster gas mass fractions have been  increased.
However, we found this solution not free of collateral effects.
We have tested this approach by running a battery of merging simulations
with  initially higher  gas fractions.

The simulations demonstrated that in an SIDM merger model it is possible to 
reproduce the observed X-ray morphology, as long as 
the initial cluster gas fractions are raised to cosmological levels 
($ f_g \sim 0.16$). 
However, as a consequence of this assumption, the final X-ray luminosity 
$L_X$ is found to be higher than the observational value 
($L_X \sim 2\cdot 10^{45} \ergs$) by a factor of $\sim 3$.

Because the bulk of the X-ray emission comes from the SE cool core, 
  the final X-ray luminosity $L_X$  can be reduced within the observational 
range by increasing the initial size of inner SE cool core region and 
consequently  reducing the cuspiness of the central gas density peak.
However, it turns out that this choice has undesirable side effects.

Specifically, we find that SIDM merger models that satisfy these initial 
conditions have  a negative final offset $d^{SE}_{X-DM} $, with  the X-ray peak 
now  trailing the DM centroid.

We explain this finding as a consequence of the larger gas core radius 
of the secondary with respect that of the SIDM model of Section \ref{sec:sidmb}, 
This implies that during its motion through the ICM of the primary,
 the secondary's cool core  will then experience 
a larger ram pressure force,  and accordingly  a larger deceleration
 \citep[see also Section 3.4 of ][]{V24}.

iv) As outlined in point i),  a  significant aspect of the SIDM merger model 
of Section \ref{sec:sidmb} 
is that the best match to the observed offsets  is obtained  from
simulations having ${\sigma_{DM}}/{m_X} $ around $ \sim 4\sxu$. 
This range of values is in tension with present  constraints 
 on  galaxy  cluster scales  \citep{Rob17,Rob19,Kim17,Shen22,Cross23}. 
For example,  upper bounds on the  galaxy-DM offset 
\citep[$\simlt 20 \kpc$, ][]{Randall08} 
were  used in SIDM merging simulations \citep{Mark04,Randall08,Rob17} 
 of the Bullet Cluster to derive upper bounds  of  approximately 
${\sigma_{DM}}/{m_X} \simlt 1 \sxu$ on the DM cross-section.

As a possible solution to this problem, 
 we proposed  \citep{V24}  that the adopted SIDM merger model 
 should be viewed as a first approximation to the physical description of DM 
 interactions. In particular, we argued that the DM collisional properties
 should be  closely related to  the collisional energy of the 
 merging cluster.

According to this view, DM interactions between the 
 two DM halos  will take place  during the collision 
as soon as  the  collisional energy $E_{\rm coll}$ of  the merging cluster 
 exceeds some critical energy threshold $E_{\rm crit}$.
For a cluster merger with  an energy  below the threshold value $E_{\rm coll}$, 
the two DM  halos will exhibit their usual  properties 
and will remain collisionless throughout  the  merging process.
According to this hypothesis, the observed offsets  should be positively 
correlated with the collisional energy  $E_{\rm coll}$ of  the merging cluster. 

For the SIDM merger model presented here, we estimate a collisional energy of 
 about $ E_{\rm{EG}} \sim 1.4 \cdot 10^{64} {\rm{\,ergs}} $ \citep{V24}.
A well-known example of a cluster merger  is the  Bullet Cluster 
\citep{Spr07,Mas08,La14,Lage15,Rob17}, for which there is no evidence of a 
significant galaxy-DM offset. 
The collisional energy  of this merging system  can be estimated to be about 
$ E_{\rm{Bullet}} \sim 3 \cdot 10^{63} {\rm{\,ergs}} $ \citep{Lage15}.
This suggests that the value of $E_{\rm crit}$  should lie between these two 
estimates: $ E_{\rm{Bullet}} \simlt E_{\rm crit}   \simlt E_{\rm{EG}} $.

This scenario will be further corroborated if other major merger clusters, as massive 
as El Gordo, are  found to  exhibit large galaxy-DM 
peak offsets ($\sim 100 \kpc$). 
 Another major merger which satisfies these constraints is the Sausage Cluster 
 CIZA J2242.8+5301 at $z=0.19$  \citep[ see  Table 2 of][]{Kim17}.

This merging system has a total  mass \citep{Jee15} of  about 
$\sim 2\cdot 10^{15} \msun$ and a mass ratio close to unity.
 The two  DM halos  are separated by about $\sim 1 \mpc$, with the  
 galaxy-DM offsets  of the order of $\sim 50-300 \kpc$ \citep{Kim17}.
  For the northern group  it is worth noting that the  DM  peak appears 
  trailing the galaxy centroid.  The  collisional energy can be estimated to 
  be approximately 
  $ E_{\rm{Sausage}} \sim 1.5 \cdot 10^{64} {\rm{\,ergs}} $  \citep{Jee15}.

 We argue that the approximate equality  $ E_{\rm{EG}} \sim E_{\rm{Sausage}} $ 
further supports  the idea that  DM behavior  in merging cluster is 
regulated by  the existence of an  energy threshold  $E_{\rm crit}$.

v) The most significant  drawback of the SIDM merger model presented here is the 
different behavior at small angles between the measured  tangential shear 
profiles and the ones extracted from the SIDM merging simulation. We now 
present a critical analysis showing the difficulty of avoiding this tension 
in the considered SIDM context.

Observationally, the binned lensing profiles of both the NW and SE clusters 
exhibit a divergent behavior as $\theta \rightarrow 0$. This  has been 
independently confirmed by several authors \citep{Jee14,Kim21} and is in accord
 with what is predicted by an NFW model to describe the halo density profile of each
 cluster.

The differences between the measured profiles and the simulated lensing 
profiles $g^{Burk}_T(\theta)$  are largest in the innermost bin,  this is because
at small angles $g^{Burk}_T(\theta)$  tends to zero.
From the size of the error bars of Figure \ref{fig:profGT}  it can be seen that
 for the SE cluster $g^{Burk}_T(\theta)$  at $\theta \sim 20^{''}$   
would still be within the $\sim 2 \sigma$ uncertainty intervals of the 
 measured tangential shear profiles. 

However, for the NW cluster this is not 
valid because of the normalization issues which affect the profiles
  $g^{Burk}_T(\theta)$  at  large angles (see the related  discussion in 
Section \ref{sec:glens}).
These disagreements can be crudely taken into account  by estimating 
at $\theta \sim 200^{''}$ the offset 
$\Delta g=g^{Burk}_T(\theta)-g^{NFW}_T(\theta)$ and  
rigidly shifting the profiles $g^{Burk}_T(\theta)$ downwards
by the corresponding amount  $\Delta g$. 
It can be easily verified that for $\theta \sim 20^{''}$  
the resulting profiles
$ \gtilde^{Burk}_T(\theta)=g^{Burk}_T(\theta)-\Delta g$ 
 are now within the $\sim 2 \sigma$ uncertainty intervals of the measured
lensing profiles of the NW cluster.

Finally, this tension may be lessened by performing for the El Gordo cluster
a SIDM merging simulation with a lower value for the SIDM cross-section, 
say ${\sigma_{DM}}/{m_X} \sim 2  \sxu$. Such a choice cannot be excluded a 
priori, but previous SIDM merger simulations (see point i) above) 
showed that merging runs with ${\sigma_{DM}}/{m_X} \sim 2 \sxu$  
are marginally inconsistent with
the observed offset $d^{SE}_{X-DM}$ \citep{V24}.

To summarize, the points discussed above lead to contradictory conclusions
regarding the SIDM merger model presented here.
Points i) and ii) being clearly in favor of an SIDM scenario for the 
merging cluster El Gordo, the interactions of DM during the collision
being able to explain the observed offsets as well
as  the magnitude of the mean relative  line-of-sight  radial velocity between 
the NW and SE clusters. About point i) it is worth noting 
that a clear benefit of an SIDM merger model for the El Gordo cluster 
is  that it can consistently  explain  all of the observed
offsets, at variance to the  results from collisionless CDM models.

The difficulties associated with points iii) to v) are of different origin 
and severity. 
Specifically,  the observed twin-tailed X-ray morphology  cannot be reproduced
faithfully by the SIDM merger model  presented  here due to  its
inaccurate modeling of the DM gravitational field during the cluster merger.
This is demonstrated by the inconsistencies raised by points iv) and v), and we 
argue that point iii) will  most likely be solved once these points 
are clarified. 

We further suggest that points iv) and v)  are closely related, 
with their inconsistencies appearing as different aspects 
of our limited knowledge about the nature of DM.
The most serious challenge faced by the proposed SIDM merger model  
is clearly that discussed in point v):
from the presented considerations, it appears that the tension 
at small angles between 
the measured  tangential shear profiles and the corresponding  profiles derived 
from the SIDM merging simulation cannot  be easily reconciled, 
at least within the given observational constraints and those derived 
from point i) on ${\sigma_{DM}}/{m_X} $.

The behavior of DM during the merger of the El Gordo cluster
is therefore somewhat contradictory: 
according to lensing data 
 internal motions of  the DM halos
are well described by a collisionless matter component,
but at the same time an SIDM merger model supports 
the presence of DM collisional properties 
as far as it concerns   the dynamic between the halos during the collision.

Our final conclusion is that, among the  alternative DM models  proposed so far 
to solve  the difficulties of the collisionless standard CDM scenario,
there is no obvious solution to this inconsistency.
On the other hand, we argue that this tension will greatly help to  unveil the 
true nature of DM by providing  a unique test bed to future theoretical DM models.


\begin{thebibliography}{99}
\bibitem{Molnar16}
 {Molnar}, S.,  \emph{Cluster Physics with Merging Galaxy Clusters}, 
\href{https://doi.org/10.3389/fspas.2015.00007}
{\emph{ Front. Astr. Space Sci.},  {\textbf 2} , (2016) 7}

\bibitem{Tulin18}
{Tulin}, S. \& {Yu}, H.-B., 
\emph{ Dark matter self-interactions and small scale structure},
\href{https://doi.org/10.1016/j.physrep.2017.11.004}
{\emph{ Phys. Reports }, {\textbf{730}},  (2018), 1}

\bibitem{Mark02}
{Markevitch}, M.,   {Gonzalez}, A.~H.,  {David}, L.,  et al., 
\emph{A Textbook Example of a Bow Shock in the Merging Galaxy Cluster 1E 0657-56}
\href{https://doi.org/10.1086/339619}
{\emph{ Astrophys. J. Lett.}, {\textbf {567}}, (2002), L27 }

\bibitem{Clowe06}
{Clowe}, D.,  {Brada{\v{c}}}, M.,  {Gonzalez}, A.~H. et al.,
\emph{A Direct Empirical Proof of the Existence of Dark Matter},
\href{https://doi.org/10.1086/508162}
{\emph{ Astrophys. J. Lett.}, {\textbf {648}}, (2006), L109 }


\bibitem{Spr07}
Springel, V.  \& {Farrar}, G.~R.,  
\href{https://doi.org/10.1111/j.1365-2966.2007.12159.x}
{\emph { MNRAS}, {\textbf {380}}, (2007), 911 }

 \bibitem{Mas08}
{Mastropietro}, C. \&  {Burkert}, A., 
\emph{ Simulating the Bullet Cluster},
\href{https://doi.org/10.1111/j.1365-2966.2008.13626.x}
{\emph { MNRAS}, {\textbf {389}}, (2008), 967 }

\bibitem{La14}
{Lage}, C. \& {Farrar}, G.~R.,  
\emph{Constrained Simulation of the Bullet Cluster},
\href{https://doi.org/10.1088/0004-637X/787/2/144}
{\emph{ Astrophys. J.}, {\textbf {787}}, (2014), 144 }



\bibitem{Donnert14}
Donnert, J.~M.~F., 
\emph{ Initial conditions for idealized clusters mergers, simulating ‘El Gordo’},
\href{https://doi.org/10.1093/mnras/stt2291}
{\emph { MNRAS}, {\textbf {438}}, (2014), 1971 }

\bibitem{Molnar15}
Molnar, S.~M. \&{Broadhurst}, T., 
\emph{A Hydrodynamical Solution for the "Twin-tailed" Colliding Galaxy 
Cluster "El Gordo"},
\href{https://doi.org/10.1088/0004-637X/800/1/37}
{\emph{ Astrophys. J.}, {\textbf 800}, (2015), 37 }

 \bibitem{Zh15}
 {Zhang}, C.,  {Yu}, Q. \&  {Lu}, Y.,  
\emph{Simulating the Galaxy Cluster “El Gordo” and Identifying the Merger 
Configuration},
\href{https://doi.org/10.1088/0004-637X/813/2/129}
{\emph{ Astrophys. J.}, {\textbf 813}, (2015), 129 }

 \bibitem{Zh18}
{Zhang}, C.,  {Yu}, Q. \&  {Lu}, Y.,  
\emph{Simulating the Galaxy Cluster “El Gordo”: Gas Motion, Kinetic 
Sunyaev-Zel’dovich Signal, and X-Ray Line Features},
\href{https://doi.org/10.3847/1538-4357/aaab4c}
{\emph{ Astrophys. J.}, {\textbf 855}, (2018), 36 }


\bibitem{Donnert17}
Donnert, J.~M.~F.,   {Beck}, A.~M., Dolag, K. \& {R{\"o}ttgering}, H.~J.~A.,
\emph{Simulations of the galaxy cluster CIZA J2242.8+5301 - I. Thermal model and
 shock properties},
\href{https://doi.org/0.1093/mnras/stx1819}
{\emph { MNRAS}, {\textbf 471}, (2017), 4587 }

\bibitem{Mol17}
Molnar, S.~M. \&{Broadhurst}, T., 
\emph{Shocks and Tides Quantified in the “Sausage” Cluster, CIZA J2242.8+5301 
Using N-body/Hydrodynamical Simulations},
\href{https://doi.org/10.3847/1538-4357/aa70a3}
{\emph{ Astrophys. J.}, {\textbf 841}, (2017), 46 }

\bibitem{Men12}
{Menanteau}, F., Hughes, J.~P., {Sif{\'o}n}, C.  {et al.}, 
\emph{ The Atacama Cosmology Telescope: ACT-CL J0102-4915 "El Gordo," a Massive 
Merging Cluster at Redshift 0.87},
\href{https://doi.org/10.1088/0004-637X/748/1/7}
{\emph{ Astrophys. J.}, {\textbf 748}, (2012), 7 }


\bibitem{Jee14}
{Jee}, M.~J., {Hughes}, J.~P., {Menanteau}, F., et al.,  
\emph{Weighing "El Gordo" with a Precision Scale: Hubble Space Telescope Weak-
lensing Analysis of the Merging Galaxy Cluster ACT-CL J0102-4915 at z = 0.87}, 
\href{https://doi.org/10.1088/0004-637X/785/1/20}
{\emph{ Astrophys. J.}, {\textbf 785}, (2014), 20 }

\bibitem{Zi13}
{Zitrin}, A., {Menanteau}, F., {Hughes}, J.~P. et al.,  
\emph{A Highly Elongated Prominent Lens at z = 0.87: First Strong-lensing 
Analysis of El Gordo},
\href{https://doi.org/10.1088/2041-8205/770/1/L15}
{\emph{ Astrophys. J. Lett.}, {\textbf {770}}, (2013), L15 }



\bibitem{As21}
{Asencio}, E.,   {Banik}, I.  \& {Kroupa}, P.,  
\emph{A massive blow for $\Lambda$CDM - the high redshift, mass, and collision 
velocity of the interacting galaxy cluster El Gordo contradicts concordance 
cosmology},
\href{https://doi.org/10.1093/mnras/staa3441}
{\emph { MNRAS}, {\textbf {500}}, (2021), 5249 }

\bibitem{As23}
{Asencio}, E.,   {Banik}, I.  \& {Kroupa}, P.,  
\emph{The El Gordo Galaxy Cluster Challenges $\Lambda$CDM for Any Plausible 
Collision Velocity},
\href{https://doi.org/10.3847/1538-4357/ace62a}
{\emph{ Astrophys. J.}, {\textbf {954}}, (2023), 162 }

\bibitem{Ng15}
{Ng}, K.~Y.,  {Dawson}, W.~A., {Wittman}, D.  et al., 
\emph{The return of the merging galaxy subclusters of El Gordo?},
\href{https://doi.org/10.1093/mnras/stv1713}
{\emph { MNRAS}, {\textbf {453}}, (2015), 1531 }

\bibitem{Molnar12}
{Molnar}, S.~M.,  {Hearn}, N.~C. \& {Stadel}, J.~ G.,  
\emph{Merging Galaxy Clusters: Offset between the Sunyaev-Zel'dovich Effect and 
X-Ray Peaks},
\href{https://doi.org/10.1088/0004-637X/748/1/45}
{\emph{ Astrophys. J.}, {\textbf {748}}, (2012), 45 }

\bibitem{Die20}
Diego, J.~M.,  {Molnar}, S.~M.,  {Cerny}, C.  et al., 
\emph{Free-form Lens Model and Mass Estimation of the High-redshift Galaxy 
Cluster ACT-CL J0102-4915, "El Gordo"},
\href{https://doi.org/10.3847/1538-4357/abbf56}
{\emph{ Astrophys. J.}, {\textbf {904}}, (2020), 106 }

\bibitem{Kim21}
Kim, J., Jee, M.~J.,  Hughes, J.~P.  et al. , 
\emph{Head-to-Toe Measurement of El Gordo: Improved Analysis of the Galaxy 
Cluster ACT-CL J0102-4915 with New Wide-field Hubble Space Telescope Imaging 
Data}
\href{https://doi.org/10.3847/1538-4357/ac294f}
{\emph{ Astrophys. J.}, {\textbf {923}}, (2021), 101 }

\bibitem{Kim17}
Kim, S.~Y., Peter, A.~H.~G. \& Wittman, D. et al., 
\emph{In the wake of dark giants: new signatures of dark matter self-interactions 
in equal-mass mergers of galaxy clusters},
\href{https://doi.org/10.1093/mnras/stx896}
{\emph { MNRAS}, {\textbf {469}}, (2017), 1414 }

\bibitem{V24}
Valdarnini, R., 
\emph{An N-body/hydrodynamical simulation study of the merging cluster El Gordo: 
A compelling case for self-interacting dark matter?},
\href{https://doi.org/10.1051/0004-6361/202348000}
{\emph{ Astr. \& Astrophys. }, {\textbf {684}}, (2024), A102 }


\bibitem{V16}
Valdarnini, R., 
\emph{Improved Performances in Subsonic Flows of an SPH Scheme with Gradients 
Estimated Using an Integral Approach},
\href{https://doi.org/10.3847/0004-637X/831/1/103}
{\emph{ Astrophys. J.}, {\textbf {831}}, (2016), 103 }

\bibitem{Du08}
Duffy, A.~R., Schaye, J., Kay, S.~T. \& Dalla Vecchia, C., 
\emph{Dark matter halo concentrations in the Wilkinson Microwave Anisotropy 
Probe year 5 cosmology},
\href{https://doi.org/10.1111/j.1745-3933.2008.00537.x}
{\emph { MNRAS}, {\textbf {390}}, (2008), L64 }

\bibitem{Bu95}
Burkert, A.,   
\emph{The Structure of Dark Matter Halos in Dwarf Galaxies},
\href{https://doi.org/10.1086/309560}
{\emph{ Astrophys. J. Lett.}, {\textbf {447}}, (1995), L25 }



\bibitem{Mer06}
  {Merritt}, D.,  {Graham}, A.~ W.,  {Moore}, B.,  {Diemand}, J.
   \& {Terzi{\'c}}, B., 
\emph{Empirical Models for Dark Matter Halos. I. Nonparametric Construction of 
Density Profiles and Comparison with Parametric Models},
\href{https://doi.org/10.1086/508988}
{\emph{ Astr. J.}, {\textbf {132}}, (2006), 2685 }

\bibitem{Kr18}
Kravtsov, A.~V., Vikhlinin, A.~A. \&{Meshcheryakov}, A.~V., 
\emph{ Stellar Mass—Halo Mass Relation and Star Formation Efficiency in 
High-Mass Halos},
\href{https://doi.org/10.1134/S1063773717120015}
{\emph{ Astr. L.}, {\textbf {44}}, (2018), 8 }

\bibitem{VS21}
Valdarnini, R. \&  {Sarazin}, C.~L., 
\emph{A study of cool core resiliency and entropy mixing in simulations of 
galaxy cluster mergers},
\href{https://doi.org/10.1093/mnras/stab1126}
{\emph { MNRAS}, {\textbf {504}}, (2021), 5409 }


\bibitem{Vog12}
Vogelsberger, M., Zavala, J. \&  Loeb, A., 
\emph{Subhaloes in self-interacting galactic dark matter haloes}, 
\href{https://doi.org/10.1111/j.1365-2966.2012.21182.x}
{\emph { MNRAS}, {\textbf {423}}, (2012), 3740 }

\bibitem{Itoh98}
Itoh, N.,  Kohyama, Y. \& Nozawa, S., 
\emph{Relativistic Corrections to the Sunyaev-Zeldovich Effect for Clusters of Galaxies},
\href{https://doi.org/10.1086/305876}
{\emph{ Astrophys. J.}, {\textbf {502}}, (1998), 7 }

\bibitem{ZuH19}
{ZuHone}, J.~A., Zavala, J. \& Vogelsberger, M., 
\emph{Sloshing of Galaxy Cluster Core Plasma in the Presence of Self-interacting 
Dark Matter},
\href{https://doi.org/10.3847/1538-4357/ab321d}
{\emph{ Astrophys. J.}, {\textbf {882}}, (2019), 119 }

\bibitem{Um20}
Umetsu, K., 
\emph{Cluster-galaxy weak lensing},
\href{https://doi.org/10.1007/s00159-020-00129-w}
{\emph { Astr. \& Astrophys. Review }, {\textbf {28}}, (2020), 7 }

\bibitem{Wri20}
Wright, C.~O.  \&  {Brainerd}, T.~G., 
\emph{Gravitational Lensing by NFW Halos},
\href{https://doi.org/10.1086/308744}
{\emph{ Astrophys. J.}, {\textbf {534}}, (2000), 34 }


\bibitem{Martel14}
{Martel}, H., {Robichaud}, F. \&  Barai, P.,  
\emph{Major Cluster Mergers and the Location of the Brightest Cluster 
Galaxy},
\href{https://doi.org/10.1088/0004-637X/786/2/79}
{\emph{ Astrophys. J.}, {\textbf {786}}, (2014), 79 }


\bibitem{Rob17}
Robertson, A.,  Massey, R. \& Eke, V.,  
\emph{What does the Bullet Cluster tell us about self-interacting dark matter?},
\href{https://doi.org/10.1093/mnras/stw2670}
{\emph { MNRAS}, {\textbf {465}}, (2017), 569 }

\bibitem{Rob19}
Robertson, A., Harvey, D., Massey, R.  et al., 
\emph{Observable tests of self-interacting dark matter in galaxy clusters: 
cosmological simulations with SIDM and baryons},
\href{https://doi.org/10.1093/mnras/stz1815}
{\emph { MNRAS}, {\textbf {488}}, (2019), 3646 }


\bibitem{Shen22}
Shen, X., Brinckmann, T.,  Rapetti, D.  et al., 
\emph{X-ray morphology of cluster-mass haloes in self-interacting dark matter},
\href{https://doi.org/10.1093/mnras/stac2376}
{\emph { MNRAS}, {\textbf {516}}, (2022), 1302 }


\bibitem{Cross23}
Cross, D., Thoron, G., Jeltema, T.~E., et al., 
\emph{Examining the self-interaction of dark matter through central cluster galaxy 
offsets},
\href{https://doi.org/10.1093/mnras/stae442}
{\emph { MNRAS}, {\textbf {529}}, (2024), 52 }

\bibitem{Randall08}
Randall, S.~W., Markevitch, M,  Clowe, D., Gonzalez, A.~ H. \&
Brada{\v{c}}, M., 
\emph{Constraints on the Self-Interaction Cross Section of Dark Matter from Numerical 
Simulations of the Merging Galaxy Cluster 1E 0657-56},
\href{https://doi.org/10.1086/587859}
{\emph{ Astrophys. J.}, {\textbf {679}}, (2008), 1173 }
 
\bibitem{Mark04}
Markevitch, M.,  Gonzalez, A.~H.,  {Clowe}, D.  et al., 
\emph{Direct Constraints on the Dark Matter Self-Interaction Cross Section from 
the Merging Galaxy Cluster 1E 0657-56},
\href{https://doi.org/10.1086/383178}
{\emph{ Astrophys. J.}, {\textbf {606}}, (2004), 819 }


\bibitem{Lage15}
{Lage}, C. \& {Farrar}, G.~R.,  
\emph{The Bullet Cluster is not a Cosmological Anomaly},
\href{https://doi.org/10.1088/1475-7516/2015/02/038}
{\emph{ JCAP}, {\textbf {2015}}, (2015) 38}

\bibitem{Jee15}
 {Jee}, M. J.  Stroe, A., Dawson, W., {et al.},  
\emph{MC 2: Constraining the Dark Matter Distribution of the Violent Merging Galaxy 
Cluster CIZA J2242.8+5301 by Piercing through the Milky Way},
\href{https://doi.org/10.1088/0004-637X/802/1/46}
{\emph{ Astrophys. J.}, {\textbf {802}}, (2015), 46 }


\end{thebibliography}

\end{document}